\documentstyle[pra,aps,12pt]{revtex}
\begin{document}
\title{Experimental Demonstration of Optimal Unambiguous State Discrimination}
\author{Roger B. M. Clarke$^{(1)}$, Anthony Chefles$^{(2)}$, Stephen M. Barnett$^{(1)}$ and Erling Riis$^{(1)}$}
\address{$^{(1)}$Department of Physics and Applied Physics, University of Strathclyde,
       Glasgow G4 0NG, UK}
\address{$^{(2)}$Department of Physical Sciences, University of Hertfordshire,
       Hatfield AL10 9AB, Herts, UK}

\input epsf
\epsfverbosetrue

\def\id{\hat{\leavevmode\hbox{\small1\kern-3.2pt\normalsize1}}}%

\maketitle
\vspace{1cm}

\begin{center}
\end{center}

\maketitle \vspace{3cm}

\begin{abstract}
\\

We present the first full demonstration of unambiguous state
discrimination between non-orthogonal quantum states. Using a
novel free space interferometer we have realised the optimum
quantum measurement scheme for two non-orthogonal states of light,
known as the Ivanovic-Dieks-Peres (IDP) measurement. We have for
the first time gained access to all three possible outcomes of
this measurement. All aspects of this generalised measurement
scheme, including its superiority over a standard von Neumann
measurement, have been demonstrated within 1.5\% of the IDP
predictions.

\end{abstract}

\setcounter{equation}{0}

\newpage

 One of the major themes of the emerging subject of quantum
information is the classical information bearing capabilities of
quantum systems.  In classical physics, different signal states
are, at least in principle, fully distinguishable, although errors
do occur due the practical difficulty of eliminating noise. By
contrast, the identification of signals carried by quantum states
will, in general, be imperfect.  This a consequence of the nature
of the quantum measurement process, which implies that only
orthogonal states can be distinguished perfectly.   Consider two
parties, Alice and Bob. Alice sends Bob a system prepared in some
quantum state chosen from a set $\{|{\psi}_{j}{\rangle}\}$ known
to Bob. Bob cannot construct a measuring apparatus which will
conclusively identify which state Alice sent with zero probability
of error unless the states $|{\psi}_{j}{\rangle}$ are orthonormal.

If the states are non-orthogonal, then Bob is forced by physical
law to weaken the specifications of his measurement.  He can relax
the condition of accuracy, in which case his measurement result
will sometimes be incorrect.  If the signal is in one of two
possible states, $|{\psi}_{\pm}{\rangle}$, with respective a
priori probabilities ${\eta}_{\pm}$, then the minimum error
probability is given by the Helstrom bound\cite{Helstrom}:
\begin{equation}
P_{e}({\mathrm
opt})=\frac{1}{2}\left(1-\sqrt{1-4{\eta}_{+}{\eta}_{-}|{\langle}{\psi}_{+}|{\psi}_{-}{\rangle}|^{2}}\right).
\end{equation}
Notice that this is zero only when the states are orthogonal.
Minimum error measurements always identify the signal as being one
of the possible states, which is to say that they are
\emph{conclusive}, although this identification will be
\emph{incorrect} with probability $P_{e}({\mathrm opt})$.  The
Helstrom measurement has recently been demonstrated in the
laboratory \cite{BRiis}, using weak optical pulses where the two
states were non-orthogonal polarisation states.

The other option available to Bob is to drop the requirement of
conclusiveness: that is, the condition that the measurement result
will always announce one of the possible states. This kind of
strategy was first described by Ivanovic\cite{Ivanovic}, who
showed that it allows two non-orthogonal states to be
discriminated \emph{without error}, but with a finite probability
of getting a third and \emph{inconclusive} result. The optimum
strategy of this kind is that which minimises the probability of
inconclusive results. Further work by Dieks\cite{Dieks} and
Peres\cite{Peres} established this minimum, which is given by the
Ivanovic-Dieks-Peres (IDP) bound:
\begin{equation}
\label{inc}
 P_{?}({\mathrm
opt})=|{\langle}{\psi}_{+}|{\psi}_{-}{\rangle}|.
\end{equation}
This bound applies when the two states appear with equal a priori
probabilities. A more general bound for two states with arbitrary
a priori probabilities was later obtained by Jaeger and
Shimony\cite{JShimony}.

Non-orthogonal photon polarisation states are also well-suited to
the realisation of this kind of measurement.  Indeed, unambiguous
discrimination between non-orthogonal polarisations in the
vicinity of the IDP limit has been carried out by Huttner \emph{et
al}\cite{Huttner}. In this experiment, linearly polarised, weak
optical pulses ($\sim$0.1 photons/pulse) were transmitted though
an optical fiber with polarisation-dependent loss. This loss was
adjusted so that the photons which were not absorbed emerged in
one of two orthogonal states, corresponding to the two
non-orthogonal input states, which were then measured in a von
Neumann measurement. Occasions when the polarisation-dependent
loss resulted in photon absorption were interpreted as
inconclusive results, and the inferred loss was in agreement with
the IDP prediction. However, the inconclusive results could not be
positively confirmed, since there were other reasons for
non-detection. Also, the experiment was only performed using one
pair of input states.

In this Letter, we report an implementation of unambiguous
polarisation discrimination at the IDP limit using free-space
interferometry, overcoming the limitations of the fiber-based
implementation. Importantly, it allows total access to all output
ports, particularly those corresponding to the three outcomes of
the IDP measurement. The absorption in our interferometer is
negligible implying that all input photons will result in either
conclusive discrimination or inconclusive results in accordance
with the IDP bound. Consequently, our experiment is the first full
demonstration of the IDP measurement.

The measurement scheme was designed to discriminate between two
non-orthogonal linear polarisation states of light. An optical
interferometer using polarising beamsplitters and waveplates was
used to physically separate appropriate polarisation components of
the input light, manipulated them, and recombine them to perform
the final measurement.  Our experiment is based on a similar
proposal by Huttner \emph{et al}\cite{Huttner}.

The states we chose to discriminate between were
\begin{equation}
|{\psi}_{\pm}{\rangle}={\cos}{\alpha}|{\leftrightarrow}{\rangle}{\pm}{\sin}{\alpha}|{\updownarrow}{\rangle},
\end{equation}

\noindent where $0{\leq}{\alpha}{\leq}45^{\circ}$, and
$|{\leftrightarrow}{\rangle}$ and $|{\updownarrow}{\rangle}$
correspond to horizontal and vertical polarisation.  These states
are depicted in figure 1.

Figure 1 also demonstrates that orthogonalisation of the two input
polarisation states, $|{\psi}_{+}{\rangle}$ and
$|{\psi}_{-}{\rangle}$, can be achieved by reducing the amplitude
along $|{\leftrightarrow}{\rangle}$.  Upon such a transformation,
they can be distinguished perfectly by a von Neumann measurement.
The remaining amplitude must not be involved in this measurement
process, and corresponds to an inconclusive result. In the
experiment of Huttner \emph{et al}\cite{Huttner} it was this
component was absorbed in the fiber cladding and could not be
measured directly. In our experiment this amplitude is detected as
light from one of the output ports of the interferometer.

The interferometer is shown in the uppermost part of Figure 2.
Isolation of the horizontal and vertical components was performed
with PBS2. PBS3 was made partially reflecting by varying the
orientation of waveplate WP4.  To orthogonalise the input states,
the amplitude of the reflected light from PBS3 must have the same
magnitude as that reflected from PBS2. The two beams are
recombined in PBS5 and analysed in a von Neumann measurement using
WP6 and PBS6. The outcome of this measurement is detected with
photodiodes PD1 and PD2. The two path lengths through the
interferometer are chosen such that input state
$|{\psi}_{+}{\rangle}$ is detected by PD1 and input state
$|{\psi}_{-}{\rangle}$ is detected by PD2. The transmission of
PBS3 is the required reduction in the horizontal component of the
input light common to both input states, and corresponds to the
inconclusive result. This light was detected on PD3 and was
defined as the loss required to orthogonalise the input states. We
measured this loss as $\alpha$ was varied from 0 to 45 degrees. It
can be seen from equation \ref{inc} that, theoretically, the
fraction of the light measured on PD3 is $\cos2\alpha$.

Although the alignment and stabilisation requirements of our
interferometer are much more stringent than for a fiber
arrangement, our apparatus offers several significant benefits.
The greatest of these is full detection of the light in the three
possible outcomes of the measurement. Indeed, it is necessary to
be able to monitor all the output ports in more complex
experiments, for example the discrimination of trine and tetrad
polarisation states\cite{Clarke}. Secondly, we are able to vary
the induced loss along the horizontal component continuously and
deterministically.

The light source was a mode-locked Ti:Sapphire laser operating at
780 nm with a repetition rate of 80.3 MHz.  The pulse duration was
1 ps, corresponding to a pulse length of 300 $\mu$m. This ensured
that there was only one pulse in the optical system at any one
time and that the length of the pulse was much shorter than the
path length of the interferometer. The output was focused with
lens L1 through a 60 $\mu$m pinhole to produce a clean spherical
wavefront. The light was then passed through lens L2 to produce a
shallow focus on the centres of the 1 mm$^{2}$ photodiodes
(Centronix, BPX65) beyond the interferometer, which were arranged
to be of equal optical path lengths away.  In this way almost
100\% of the light input into the interferometer reached the
detectors.

Neutral density filters were inserted to attenuate the light to an
average of 0.2 photons per pulse (4 pW) at the interferometer
input. Fine adjustment of the intensity was possible by rotating
the waveplate WP1, placed before the Rochon polarising
beamsplitter, PBS1.  This type of beamsplitter was chosen for its
high extinction ratio, measured to be greater than 1 part in 5000.
A Wollaston type beamsplitter, PBS6, was used in the analyser part
of the experiment for the same reason.

The input to the interferometer was obtained from the linearly
polarised straight through beam of PBS1. The polarisation state of
this input beam was changed using two half waveplates, WP2 and
WP3.  The first rotated the polarisation angle from 0 to 45
degrees above the horizontal, preparing the polarisation state
$|{\psi}_{+}{\rangle}$.  The second waveplate, WP3, was oriented
such that it had the effect of flipping the polarisation state
about the horizontal axis, transforming the input state
$|{\psi}_{+}{\rangle}$ to $|{\psi}_{-}{\rangle}$.  In this way,
the input states were easily exchanged.

Due to the very low light levels, phase sensitive detection, using
a chopper wheel, was required to recover the analogue signals from
detectors 1-5. The photodiodes had a nominal quantum efficiency of
83\% and were terminated by 10 M$\Omega$, and were biased in
parallel by a single 9 V source. For light levels equivalent to
0.1 photons per pulse, the average current obtained is 1.1 pA,
which equates to 11 $\mu$V across the termination.  Regular
measurements were taken of the small offsets that arose when using
this technique. Light levels equivalent to less than 0.01 photons
per pulse were detectable.

To normalise out the amplitude noise of the Ti:Sapphire laser, a
pick-off beam was measured on PD6 when \emph{any} measurements of
photodiodes 1-5 were taken.  Again, phase-sensitive detection was
performed, but using a separate lock-in amplifier. Photodiodes 1-5
were calibrated relative to each other to better than 1\% by
changing the distribution of light around the interferometer with
waveplates.

The interferometer itself was constructed from four AR coated
polarising beamsplitters PBS2-5 mounted on a machined monolithic
aluminium block.  The optical pathlength difference in the two
arms was inferred to be less than 4 $\mu$m over the 80 mm total
pathlength from the extinction ratio obtained when used in a
Mach-Zehnder operation.  PBS5 was capable of being rotated around
and translated along all axes with piezo-electric transducers. The
AR coated ${\lambda}/2$ waveplates used were measured to maintain
the linearity of polarisation to 1 part in 2000.

The fringe visibility of the interferometer when used in a
conventional Mach-Zehnder operation was measured to be better than
200:1.  The translational and angular stability of the
interferometer was inferred to be less than 100 nm and 0.001
degrees respectively over at least half an hour. This level of
stability was vital when taking results over the prolonged periods
of time needed when using phase sensitive detection.

Calibration of the beamsplitters was performed to determine their
polarisation properties. This was particularly important for
understanding the results for small $\alpha$.  A small amount of
birefringence of the beamsplitters meant that for horizontally
polarised input light, the transmitted power was comprised of
98.2\% horizontally and 0.9\% vertically polarised light. The
remaining power was reflected and almost equally distributed
between the two polarisation components. Approximately the same
leakages were found for vertically polarised input light.

To align the interferometer the input was set to
$|{\psi}_{+}{\rangle}$, $\alpha=45$ degrees, and WP4 set so that
PBS3 reflected all the light (zero loss ideally). This resulted in
equal amplitudes of light reaching PBS5 from each arm,
corresponding to a conventional Mach-Zehnder interferometer. The
path lengths and angles were varied using PBS5 to obtain the
maximum visibility of the interference. When this occurred, the
signal from PD2 was at a minimum.

The experiment was performed by preparing the input state
$|{\psi}_{+}{\rangle}$ in approximately 4 degrees steps from
$0{\leq}{\alpha}{\leq}45$ degrees. The exact angle was inferred by
measuring the light in each arm immediately after PBS2 using all
the detectors. High input powers were used to obtain a good signal
to noise ratio.  The light transmitted through PBS3 was then
varied by rotating WP4 such that the signal from PD2 was at a
minimum. The measured signal at PD2 is defined as the measured
error rate since in theory 100\% of the light should reach PD1. In
practice, it is non-zero due to experimental imperfections.

The fraction of transmitted light through PBS3 was measured
classically as the fraction of the total power incident on PD3 and
the sum of the outputs of all the other photodiodes.

The transmittance of PBS3 was varied, using WP4, to obtain the
minimum signal on PD2 at high light levels.  It was verified that
this minimum was obtained using the same angle of WP4 at any light
level, including 0.2 photons per pulse.  High light levels were
therefore used to ensure a high signal to noise ratio in PD2 to
obtain the best measure of the transmittance of PBS3.

The intensity was then reduced to 0.2 photons per pulse and the
error rate, on PD2, measured. Given the extremely low light levels
in PD2 (typically 0.01 photons per pulse or 0.2 pW), the
measurement error of $\pm2.5$\% of the total light detected was
comparable to the signal expected at this port. Therefore the
extinction ratio of 1 in 200 could not be observed at these power
levels.  For $\alpha<15^{\circ}$, the amplitude of the light
entering the interferometer was increased up to a maximum of 1
photon per pulse in order that an error rate could be measured.

The input state was then changed from $|{\psi}_{+}{\rangle}$ to
$|{\psi}_{-}{\rangle}$ by inserting WP3. The error rate, the
signal measured on PD1 this time, was measured with no other
alteration to the apparatus. The alignment of the interferometer
was checked after changing the angle ${\alpha}$ for the next pair
of states $|{\psi}_{+}{\rangle}$ and $|{\psi}_{-}{\rangle}$.

Figure 3 shows the light on PD3, the induced loss, needed to
produce the minimum error rate of distinguishment against
$\alpha$, for $0<{\alpha}<45$ degrees. The RMS deviation from the
ideal theoretical curve is 1.3\%, clearly validating the
Ivanovic-Dieks-Peres measurement scheme. The error in PD3 was
estimated by rotating WP4 until the signal on PD1 increased
noticeably. For angles less than 30 degrees the sensitivity became
so great that this error could not be estimated quantitatively and
the uncertainty is less than the size of the points in the figure.
A model based upon the measured characteristics of the PBS's,
described in the next paragraph, reproduces the experimental
results within the estimated error values.

The error rates obtained with incident angle are plotted in figure
4.  The states $|{\psi}_{+}{\rangle}$ and $|{\psi}_{-}{\rangle}$
are shown at the same angle position.  Also plotted is error rate
associated with the best possible von Neumann measurement (Eq. 1
with $\eta_1=\eta_2=\frac{1}{2}$).  Our data clearly shows error
probabilities that are below this level.  For
$14^{\circ}<\alpha<45^{\circ}$ the average experimental error rate
for the two input states is 2.8\%. For smaller angles the error
rate rises significantly.  We modelled the behaviour of the
interferometer using non-ideal PBS's based on the calibration data
obtained previously (no phase information was available and we
assumed 0 and 90 degrees phase changes upon transmission and
reflection respectively). The experimental procedure was followed,
optimising the loss for state $|{\psi}_{+}{\rangle}$ and then
flipping to state $|{\psi}_{-}{\rangle}$ to obtain the error
rates. These results are also shown in Figure 4 and are in good
qualitative agreement with the experimental results.  For small
$\alpha$, the leakages of the PBS's are such that the errors
present are of comparable size to the ideal signals.

We have clearly demonstrated the IDP measurement scheme using a
free space interferometer.  For the first time the loss required
to obtain unambiguous state discrimination was confirmed by direct
measurement and found to be consistent with the ideal theoretical
values at the 1\% level. The low light levels used, typically 0.2
photons per pulse, were the limiting factor in measuring the error
rate for $\alpha>14^{\circ}$.  We have shown that for angles
smaller than this the performance of the polarising beamsplitters
is the limiting factor.

This work was funded by the UK Engineering and Physical Sciences
Research Council.

\newpage

\begin{figure}

\epsfxsize16cm \centerline{\epsfbox{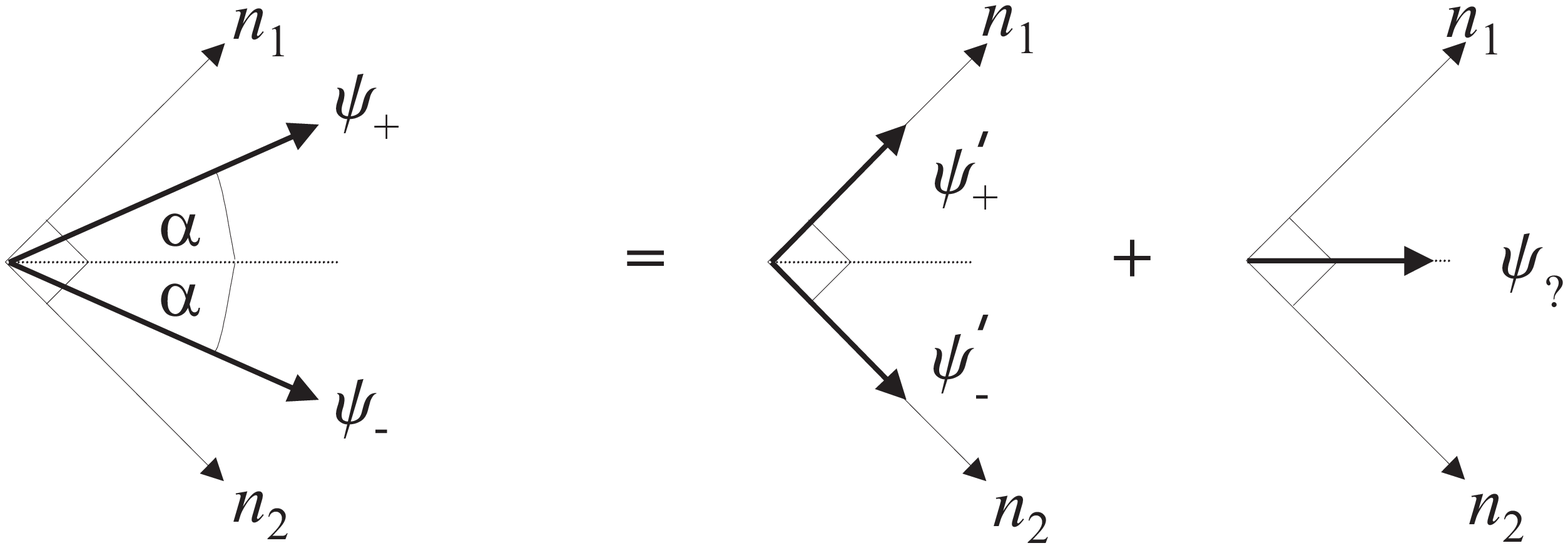}} \vspace{1cm}
\caption{The components of the states $|{\psi}_{+}{\rangle}$ and
$|{\psi}_{-}{\rangle}$ can be separated to form two auxiliary
states, $|{\psi}_{+}^{'}{\rangle}$ and $|{\psi}_{+}^{'}{\rangle}$,
and a common state $|{\psi}_{?}{\rangle}$.
$|{\psi}_{+}^{'}{\rangle}$ and $|{\psi}_{-}^{'}{\rangle}$ can be
discriminated perfectly by a von Neumann measurement along the
orthogonal basis vectors $n_{1}$ and $n_{2}$.
$|{\psi}_{?}{\rangle}$ is common to both initial states and
corresponds to an inconclusive result.}

\end{figure}

\newpage

\begin{figure}[hbt]
\label{interferometer}
 \epsfxsize=13cm
\centerline{\epsfbox{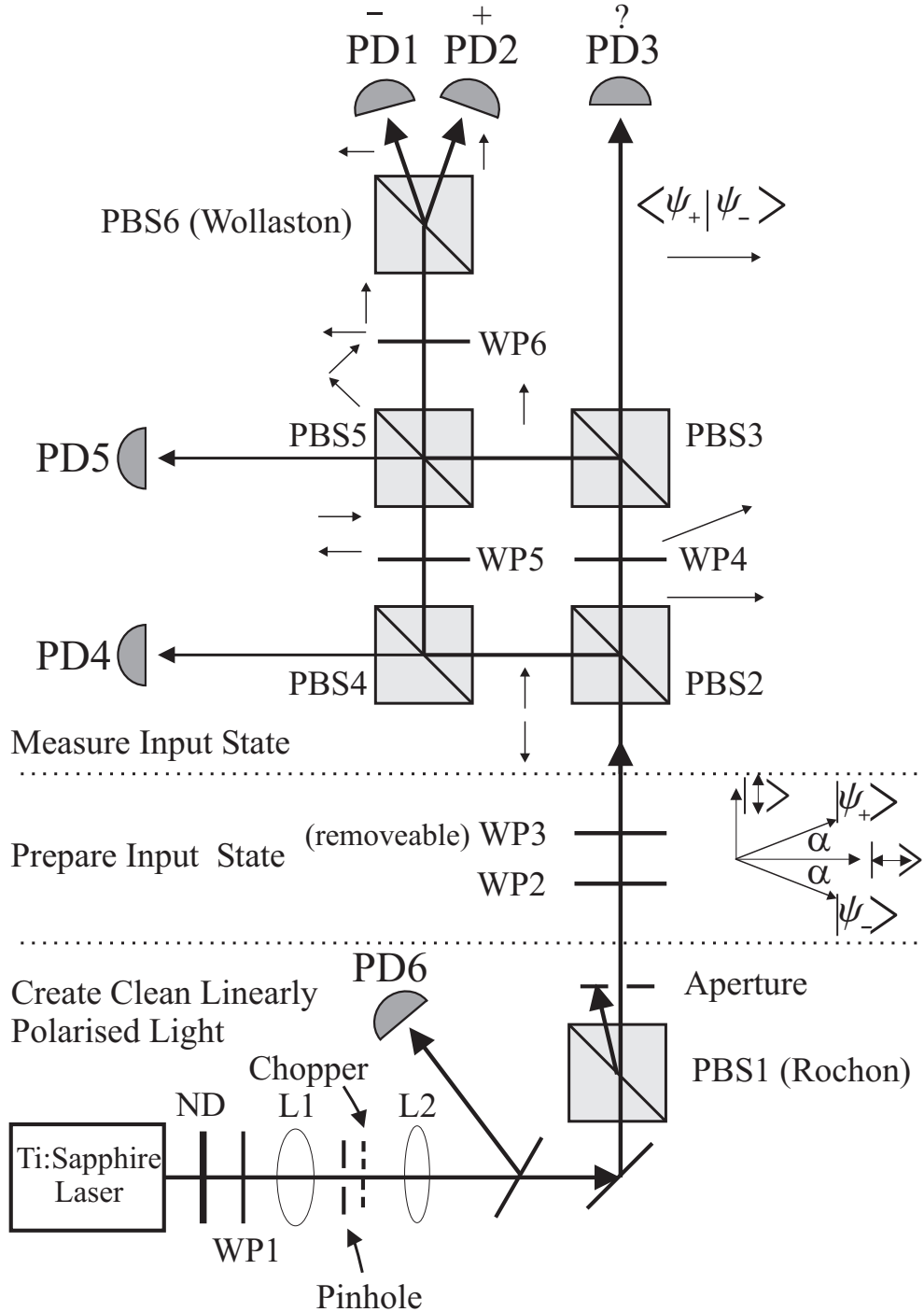}}

\vspace*{1cm}

\caption{Experimental setup to prepare, process and discriminate
the states $|{\psi}_{+}{\rangle}$ and $|{\psi}_{-}{\rangle}$.
    See text for full description.  L = lens, ND =
    neutral density filter to attenuate the light,
    PD = photodiode,
    WP = waveplate, all $\lambda$/2,
    to rotate the polarisation of the light.
    PBS = polarising beam splitter, all reflect vertical
    polarisation and transmit horizontal polarisation.}

\end{figure}

\newpage

\begin{figure}

\epsfxsize=16cm

\centerline{\epsffile{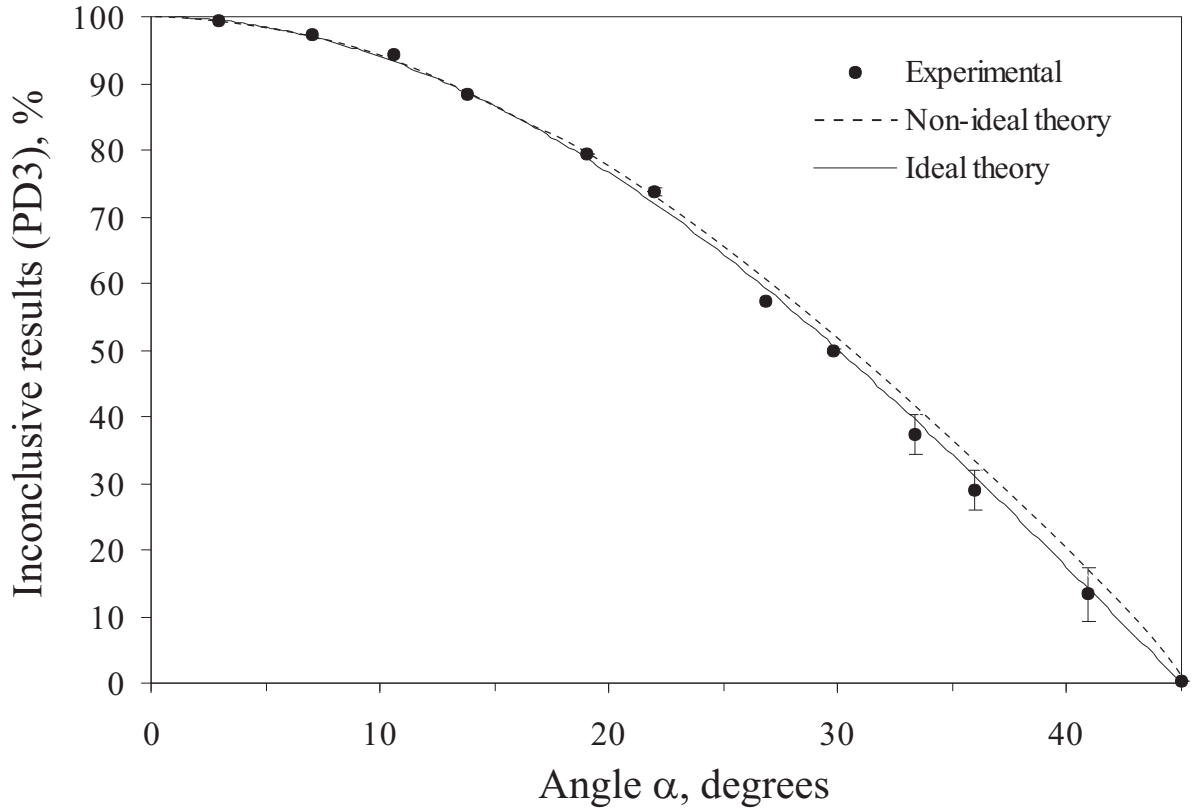}} \vspace{1cm}

\caption{The dark points show experimental signal detected on PD3,
corresponding to inconclusive results, required to orthogonalise
the input states. The errors are derived from the maximum possible
deviation required to observe a significant increase in the
discrimination error rate.  The continuous curve shows the ideal
theoretical values for the inconclusive loss rates, $\cos2\alpha$.
A model using the characteristics of the non-ideal beamsplitters
was used to generate the dashed curve. The estimated error of this
line is approximately twice the experimental error, and was
derived by optimising the extinguishment of the signal on PD2 to
the 0.1\% level.}

\end{figure}

\newpage

\begin{figure}

\epsfxsize=16cm

\centerline{\epsffile{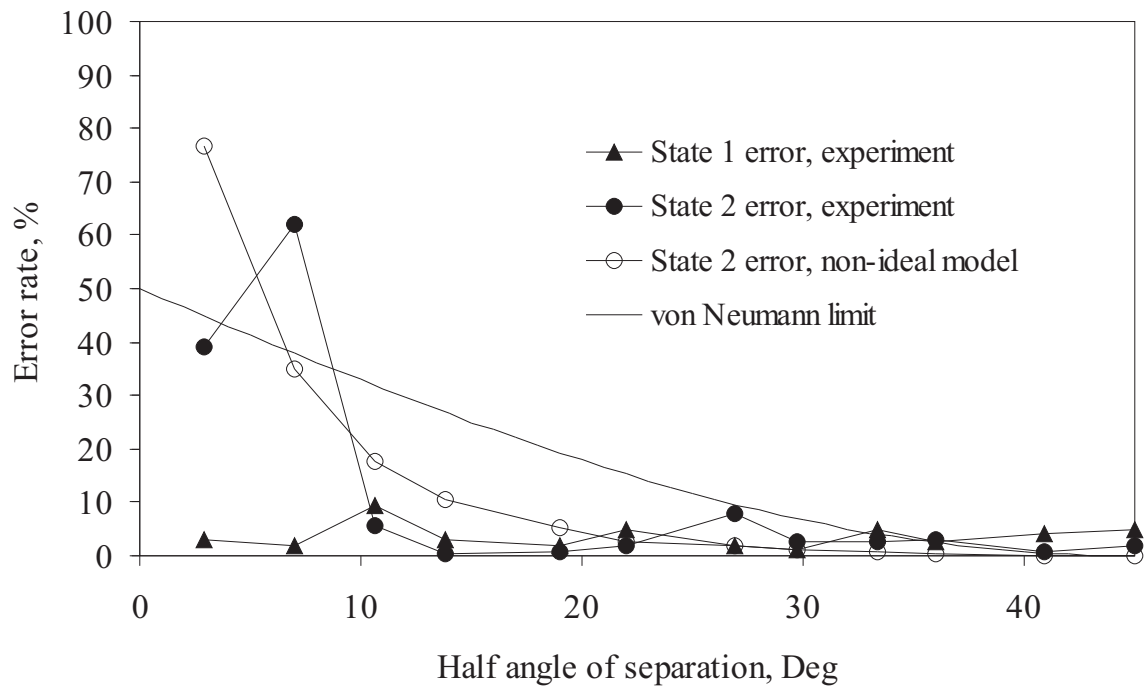}}

\vspace{1cm}

\caption{The experimentally observed error rates (PD2 and PD1)
obtained with input states $|{\psi}_{+}{\rangle}$ and
$|{\psi}_{-}{\rangle}$ respectively. Also shown is the theoretical
model using non-ideal beamsplitters with the same characteristics
as in the experiment, and the best possible von Neumann error
rate.}

\end{figure}

\end{document}